# Augmented Coaching Ecosystem for Non-obtrusive Adaptive Personalized Elderly Care on the Basis of Cloud-Fog-Dew Computing Paradigm

Yu.Gordienko[1*], S.Stirenko[1], O.Alienin[1], K.Skala[2], Z.Soyat[2], A.Rojbi[3], J.R.López Benito[4], E.Artetxe González[4], U.Lushchyk[5], L.Sajn[6], A.Llorente Coto[7], G.Jervan[8]

[1] National Technical University of Ukraine "Igor SIkorsky Kyiv Polytechic Institute" (NTUU KPI), Kyiv, Ukraine
[2] Ruder Boskovic Institute, Zagreb, Croatia
[3] University of Paris 8, Paris, France
[4] CreativiTIC Innova SL, Logroño, Spain
[5] Medical Research Center "Veritas", Kyiv, Ukraine
[6] University of Ljubljana, Ljubljana, Slovenia
[7] Private Planet, London, United Kingdom
[8] Tallinn University of Technology, Tallinn, Estonia
[*] yuri.gordienko@gmail.com

**Abstract** - The concept of the augmented coaching ecosystem for non-obtrusive adaptive personalized elderly care is proposed on the basis of the integration of new and available ICT approaches. They include multimodal user interface (MMUI), augmented reality (AR), machine learning (ML), Internet of Things (IoT), and machine-to-machine (M2M) interactions. The ecosystem is based on the Cloud-Fog-Dew computing paradigm services, providing a full symbiosis by integrating the whole range from low level sensors up to high level services using integration efficiency inherent in synergistic use of applied technologies. Inside of this ecosystem, all of them are encapsulated in the following network layers: Dew, Fog, and Cloud computing layer. Instead of the "spaghetti connections", "mosaic of buttons", "puzzles of output data", etc., the proposed ecosystem provides the strict division in the following dataflow channels: consumer interaction channel, machine interaction channel, and caregiver interaction channel. This concept allows to decrease the physical, cognitive, and mental load on elderly care stakeholders by decreasing the secondary human-to-human (H2H), human-to-machine (H2M), and machine-to-human (M2H) interactions in favor of M2M interactions and distributed Dew Computing services environment. It allows to apply this non-obtrusive augmented reality ecosystem for effective personalized elderly care to preserve their physical, cognitive, mental and social well-being.

## I. Introduction

### A. Background

The advances in medicine and living standards in the last century have resulted in a significant increase in the number of elderly people in Europe and most other developed countries in the world. Over the next decades, the worldwide number of older people will further increase dramatically. In Europe, this development is even more pronounced: for example, in Portugal, Spain, Croatia and other European countries, the old age dependency ratio, which gives the quotient of people 65+ will reach ~30-36% with pan-European average value up to 29.6% in 2050 [1]. These demographic changes have drastic structural, societal and economic implications, and challenge elderly care stakeholders like policymakers, families, businesses and healthcare providers alike. The ever increasing percentage of old people in the most advanced Western and Eastern countries is posing a great challenge in social healthcare systems. The effort required by formal caregivers for supporting older people can be enormous, and this requires an increase in the efficiency and effectiveness of today care. One way for achieving such a goal is the use of information and communication technologies (ICTs) for supporting and assisting people in their own homes.

Older generations need to be included as active and integral pillars of our society instead of being isolated in the special elderly care facilities. They should remain active members of the work force as long as possible, since the traditional assumption that retirement equals the worker's final exit from the labor force does not hold true any longer. The required transition of society can only be successful if huge efforts are made on various levels to foster independence of this age group, from more flexible employment arrangements, remote services in care giving (telecare), support of independent living (ambient assisted living - AAL), access to information, access to transportation (accessibility), to specific communication services and devices as well as entrepreneur approaches in educational offers like life-long learning (LLL).

ICT is believed to play a key role in all these fields. However, ICT can successfully contribute to their individual well being, and help to meet the challenges of an aging society in general, if ICT could be non-obtrusively adapted to the older adults' knowledge, needs, and abilities. Furthermore, the whole of our society can gain enormous benefits by integrating the knowledge and skills and high degree of experience the elderly can provide to the coming generations, in all aspects of living, from technological expertise in any field, to everyday living experiences. Current ICTs range from systems for reminding appointments and activities [2], for medical assistance and tele-healthcare [3], to human-computer interfaces for older persons or people with special needs [2]. Usually, these ICTs incorporate application dependent sensors, such as sensors, cameras or microphones. Many studies [4] have demonstrated that people prefer non-invasive sensors, such as microphones, over cameras and wearable sensors, and this drove the scientific community to develop systems and technologies based on non-invasive approaches only.

ICTs have become an integral component of everyone's life, including older adults, to continue education, obtain health information, communicate and exchange experiences, as well as online banking/shopping etc. Though recent research has shown that older adults are receptive to using ICTs, a commonly held belief is still prevalent that supports the idea that older adults are unwilling to use ICTs due to bodily and cognitive decline in working memory, attention, and spatial abilities [5,6].

The main problem is that despite the current progress of elderly care facilities the vast majority of EU older people wish to live independently at home as long as possible; meeting their needs can be a major challenge [7]. The different providers often work under conditions of poor coordination among ICT experts, elderly caregivers, patients, and their families [8-9].

*B. State of the Art (Similar Works)*

ICTs are promising for the long-term care of elderly people. As all European member states are facing an increasing complexity of health and social care, good practices in ICTs should be identified and evaluated. Recently, several projects funded by DG CNECT were related to Active and Healthy Ageing (AHA). They provided: independent living and integrated services — BeyondSilos (http://beyondsilos.eu), integrated care coordination, patient empowerment and home support — CareWell (www.carewell-project.eu), set of standard functional specifications for an ICT platform enabling the delivery of integrated care to older patients — SmartCare (http://pilotsmartcare.eu/). Some successful initiatives were initiated in Europe and supported by EU, for example, European Rosetta project [11], research network for design of environments for ageing (GAL) [10], assisted living environment for independent care and health monitoring of the elderly (ENRICHME), responsive engagement of the elderly promoting activity and customized healthcare (REACH), digital environment for cognitive inclusion (DECI), integrated intelligent home environment for the provision of health, nutrition and mobility services to the elderly (MOBISERV), unobtrusive smart environments for independent living (USEFIL), open architecture for accessible services integration and standardization (OASIS) and others.

*C. Unresolved Problems*

These innovations can improve health outcomes, quality of life and efficiency of care processes, while supporting independent living. However, in the face of new challenges some disruptive innovations should be proposed and implemented, and the new challenges/problems should be addressed. The potential radically new solution should take into account the following additional set of aspects/problems related to quite different (1) targeted communities; (2) level of functional (technical/computer/digital) literacy of the targeted communities; (3) realistic time of massive implementation of the proposed technologies for these communities with people of various functional literacy; (4) differences in national and geographical mentality as to elderly care in Europe.

Targeted communities in the context of elderly care consist of:

- individuals — self-directed elderly care, where elders control both the objectives and means of elderly care;

- families, i.e. individuals inside family and/or supported by family — informal elderly care, where elders control the means/tools, but not the objectives of elderly care;

- assisted elderly care — non-formal elderly care, where elders control the objectives but not the means/tools of elderly care;

- specialized elderly care facilities — formal elderly care, where elders have no or little control over the objectives or means/tools of elderly care.

Level of functional/computer/digital literacy of the targeted communities (in the order from the lowest to highest): absolute computer illiteracy, digital phobic, basic computer literacy, digital immigrants [12], intermediate computer literacy, digital visitors [13], proficient computer literacy, digital residents [13], digitally native [12].

The proposed time of massive implementation of the proposed technologies/environments depends on the maturity of the available solutions and the functional/computer/digital literacy level of the targeted community: now (the current mature technologies can be applied immediately), in the nearest future (the perspective technologies can be mature in the nearest 2-3 years), in the much later future (the perspective technologies can be mature at unknown time).

Differences in national and geographical mentality as to elderly care in Europe were observed and reported elsewhere [14,15]:

- informal care is more common in South than in North Europe;

- informal care is more common in the "new" member states in the "East" than in the "old" member states in the "West";

- informal care provision to someone outside the household is comparatively rare in the Mediterranean countries, elderly care to someone in the home is more common in these countries than in the EU-states on average;

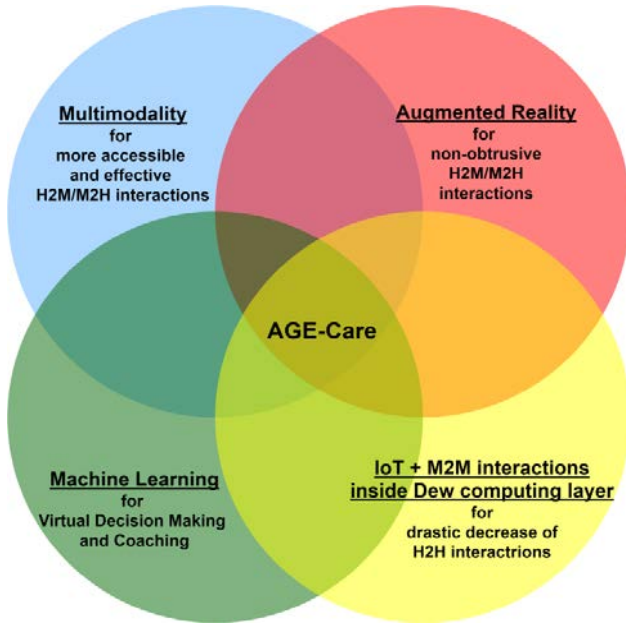

Figure 1. The integration concept of non-obtrusive augmented reality learning and coaching ecosystem for effective personalized elderly care.

- the low proportion of people providing care within households is explained by the rarity of multigenerational households in Nordic Europe.

The proposed Augmented Coaching Ecosystem for Non-obtrusive Adaptive Personalized Elderly Care (AGE-Care) is focused on the provision of the virtual care, support, and coaching to elderly people in the various targeted communities and with different functional/computer/digital literacy of the targeted communities. It will be achieved by enhancement of available ICT-enabled elderly care services, development of new ones, and their application with the tight coordination, monitoring, self-management and caregivers involvement inside the proposed AGE-Care ecosystem.

II. CONCEPT, MAIN AIMS, AND BASIC PRINCIPLES

A. General concept

The proposed AGE-Care ecosystem is assumed to be based on the integration of the several new ICT approaches and available ones, which should be enhanced by the radically new ICT based technologies concepts (shown in Figure 1) in favor of the elderly care stakeholders. They include multimodal user interface (MMUI), augmented reality (AR), machine learning (ML), Internet of Things (IoT), Internet of Everything (IoT), machine-to-machine (M2M) interactions, based on the Cloud-Fog-Dew computing paradigm services, providing a full symbiosis by integrating the whole range from low level sensors up to high level services using integration efficiency inherent in synergistic use of applied technologies.

The AGE-Care ecosystem is assumed to penetrate any organizational, national, mental, gender, and cultural division lines, boundaries, and limits. It will use the most appropriate available resources and elderly care, healthcare, and social care services. The AGE-Care ecosystem will be based on open standards, multi-vendor interoperability, collaboration with ICT suppliers and ICT-related service providers.

B. Main aims

The main aims of AGE-Care ecosystem are as follows:

- to develop, test, and validate radically new ICT based concept of non-obtrusive augmented reality learning and coaching ecosystem for effective personalized elderly care to improve and maintain their independence, functional capacity, health status as well as preserving their physical, cognitive, mental and social well-being,

- to develop and implement the synergetic user-centered design of intuitive human-to-machine (H2M) and machine-to-human (M2H) interactions on the basis of information and communication technologies (ICTs) including internet of things (IoT), multimodal augmented reality (AR), and predictive machine learning (ML) approaches,

- to decrease the physical, cognitive, and mental load on elderly care stakeholders by decreasing the secondary human-to-human (H2H), human-to-machine (H2M), and machine-to-human (M2H) interactions in favor of machine-to-machine (M2M) interactions and distributed Dew Computing services environment,

- to overcome cognitive, mental, institutional, regional, and national barriers enabling delivery of integrated elderly care on the European scale by joining efforts across governmental, non-governmental, and volunteer elderly care organizations and individuals.

The following radically new ICT based main concepts and approaches are planned to be used to reach these aims (Figure 1):

- multimodal user interface (MMUI) — for the more accessible and effective intuitive H2M/M2H interaction on the basis combination of creative "artistic" approaches;

- augmented reality (AR) — for non-obtrusive H2M/M2H interactions,

- machine learning (ML) — for virtual decision making and virtual guidance of users,

- Internet of Things (IoT) + Internet of Everything (IoT) + machine-to-machine (M2M) interactions encapsulated inside Dew computing layer — to hide "behind the curtains" the mental and cognitive overloads, and shift them from H2H to M2M interaction zone.

C. Basic Principles

The proposed open AGE-Care ecosystem is based on the several basic principles:

- dominance of machine-to-machine (M2M) interaction over human-to-human (H2H);

- multimodal instead of single-modal interactions;

- non-obtrusive augmented reality feedback instead of obtrusive direct communication with numerous high-tech sensors, actuators, devices, and gadgets;
- virtual decision making and coaching by machine learning instead of real human-related services,
- short adaptive learning curve by selection of specific and context-related virtual coaching methods based on LLL principles instead of the obsolete and awkward "user guide" and "context help" approaches;
- highly distributed service oriented local and distance communication and service facilities.

III. STRUCTURE, WORKFLOWS, AND SOME EXAMPLES

A. *Hierarhical Structure*

This basic hierarchical structure of the AGE-Care ecosystem is virtualized at different levels and visually presented in Figure 2. In contrast to the current concept of elderly care (Fig. 2a), the proposed concept (Fig. 2b) will allow stakeholders:

- to decrease significantly (and avoid in the most situations) the level of H2H interactions — by emphasis on the M2M interactions for the basic technological scenarios;
- to avoid technological H2H interactions, but emphasize emotional H2H interactions in favor of emotional positive feedback from elderly people due to involvement of augmented multimedia channels like observed and even performed art, music, dance, etc.;
- to increase efficiency of H2M/M2H interactions — by introduction of multimodal communication channels like audio, visual, tactile, odor, etc., so-called Augmented Reality Human-to-IoT (ARH2IoT) interactions;
- to increase the acceptance level of the available ICT technologies for elderly care — by providing their functional abilities through non-obtrusive augmented reality pathways;
- to eliminate the gap between the newest available ICT technologies for elderly care and computer literacy of the targeted communities — by context-related, problem-based, and personalized virtual AR-related coaching;
- to decrease the market entry threshold for the future ICT technologies for elderly care — by providing the related open platform specifications based on the best practices and lessons learned during the project;
- to provide more security and privacy — by the localization of the personal consumer data at the lower scales of the AGE-Care ecosystem.

B. *Workflows and Network Layers*

Inside of AGE-Care ecosystem all workflows are encapsulated in the following network layers:

- Dew computing layer: the raw sensor data and basic multimodal actuator actions are concentrated, pre-processed, and resumed in the smallest scale local network (Dew) at the level of the IoT-controllers (individuals) and shared with the upper Fog computing layer;
- Fog computing layer: the resumed IoT-controller data and advanced actuator actions are located in the medium scale regional network unit (Fog) at the level of the IoT-gateway (family/room/office) and shared with the lower Dew computing layer and upper Cloud computing layer;
- Cloud computing layer: the accumulated IoT-

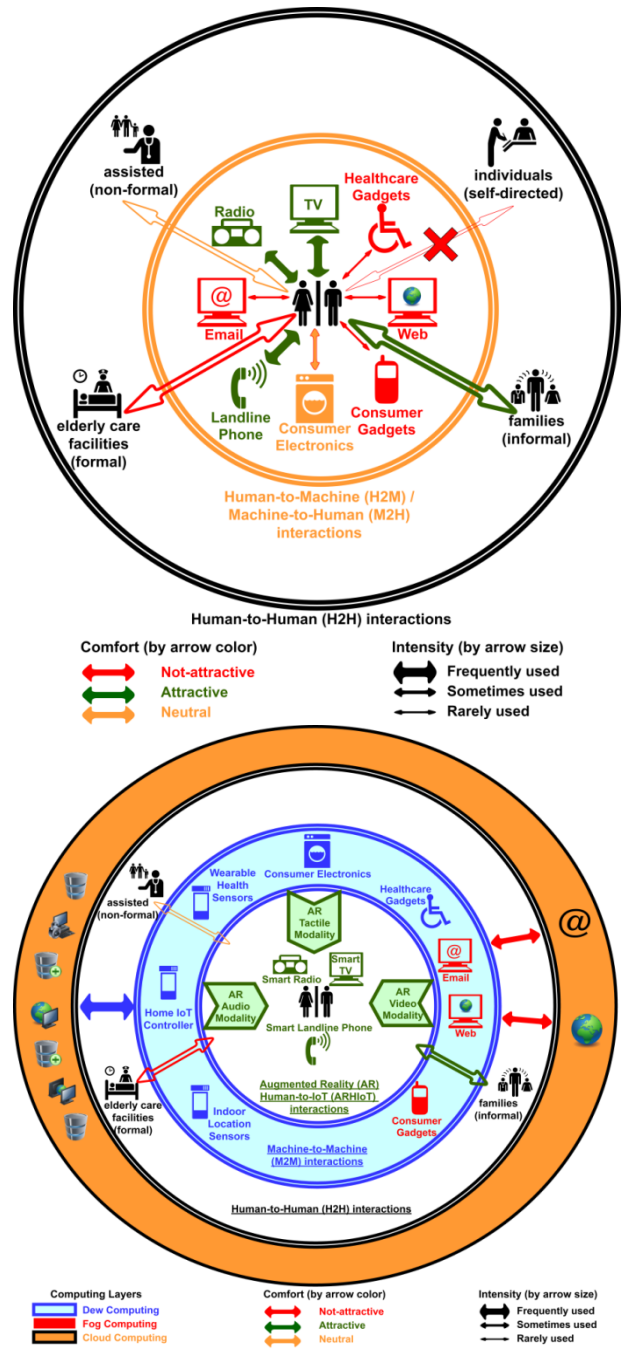

Figure 2. The current concept of elderly care (a, top), and the proposed concept of Augmented Coaching Ecosystem for Non-obtrusive Adaptive Personalized Elderly Care (AGE-Care) (b, bottom).

gateway data are thoroughly analyzed by ML methods to provide virtual decisions and coaching advices in the highest scale global network (Cloud) at the level of the global computing centers (hospitals, healthcare authorities, associations, corporations, etc.) and delivered to the lower Fog and Dew Computing layer.

*C. Communication Flows and Interactions*

The typical communication flows inside the AGE-Care ecosystem are schematically shown in Fig. 2b by arrows, where the higher emphases (in contrast to the current concept of elderly care) are placed on:

- ARH2IoT interactions — under Dew computing layer: green arrows depict the main dataflows from/to consumers by the familiar communication channels and devices, but with context-sensitive information provided by the multimodal augmented reality;
- M2M interactions — mainly inside Dew computing layer: light blue circle depicts the undercover dataflows among sensors and actuators, which are laid in the base of the multimodal augmented reality in ARH2IoT interactions;
- Cloud-Fog interactions — between Cloud and Fog computing layers: red arrows denote the familiar dataflows between the global computing centers and the medium scale network unit (Fog) at the level of the IoT-gateway (family/room/office);
- Cloud-Dew interactions — between Cloud and Dew computing layers: blue arrow denotes the dataflows between the global computing centers and the IoT-controllers.

It will allow to decrease cognitive overload on the stakeholders, because in the current concept of elderly care (Fig. 2a) the stakeholders are overwhelmed by the everyday increasing variety of the newest ICT technologies, the related devices and unusual practices. In the current paradigm of eHealth and elderly care, the stakeholders have to go by the long, complicated, and non-familiar learning curve to leverage the new ICT technologies. In contrast to it, the AGE-Care ecosystem proposes them to use the familiar information pathways (devices like television and radio broadcasting, landline phone communication), that seem to be the same old things, but actually enhanced by newest AR and AI technologies under the hood.

Instead of the "spaghetti connections" to the numerous sensors, actuators, devices, and gadgets with sporadic dataflows, "mosaic of buttons", and "puzzles of output data" for each device/technology, etc. (Fig. 2a), the AGE-Care ecosystem will provide the strict division in the following dataflow channels (Fig. 2b):

- consumer interaction channel — by allowing feedback data from all applied eHealth and elderly care ICT technologies through augmented reality pathway only at ARH2IoT layer;
- machine interaction channel — by integration of all sensor/actuator technologies and isolation of their raw data at Dew computing layer,
- caregiver interaction channel — by integration of Dew, Fog, and Cloud computing layers.

In general, the AGE-Care ecosystem will decrease the high cognitive load on customers, increase the efficiency of caregivers, and provide a unified way for incorporation of any future ICTs by division of dataflows into the above mentioned consumer, machine, and caregiver channels. This work will include the necessary formalization procedures: standardization, definitions of customer and stakeholder interfaces, identification of data models and data processing tools, and privacy and security policies and recommendations.

The necessary conditions for incorporation of the available and future ICTs to the AGE-Care ecosystem are mostly related with adaptation to the paradigms of:

- multimodal augmented reality (AR) data output for consumers;
- Dew computing (and available M2M standards inside it) for basic and automatic decision making;
- multilayer interaction between Cloud, Fog, and Dew computing for advanced (mostly automatic and limited manual) decision making.

*D. Some Implemented Combinations of Components*

Several combinations of the new ICTs (which are actually the components of the AGE-Care ecosystem) are already implemented by authors and their detailed explanation and related background can be found elsewhere in the related publications, for example:

- *Frameworks for Integration of Workflows and Distributed Computing Resources*: gateway approaches in science and education [16-18];
- *Dew (+ Fog + Cloud) computing + IoT + IoE*: the conceptual approach for organization of the vertical hierarchical links between the scalable distributed computing paradigms: Cloud Computing, Fog Computing and Dew Computing, which decrease the cost and improve the performance, particularly for IoT and IoE [19];
- *AR + visual + tactile interaction modes*: to provide tactile metaphors in education to help students in memorizing the learning terms by the sense of touch in addition to the AR tools [20,21];
- *ML + visual + tactile interaction mode*: to produce the tactile map for people with visual impairment and recognize text within the image by advanced image processing and ML [22];
- *IoT for eHealth (wearable electronics) + ML + AR + brain-computing interface + visual interaction mode*: to monitor, analyze, and estimate the accumulated fatigue by various gadgets and visualize the output data by AR means [23-25].

## IV. Conclusions

The proposed integrated ecosystem provides the basis for effective personalized elderly care by introduction of multimodal personalized communication channels. It allows end users to get cumulative effect from mixture of ICTs like IoT/IoE, multimodal AR, and predictive ML approaches. As a result, it could exclude obtrusive H2M/M2H technological interactions by delivering them to M2M interactions encapsulated in Dew Computing layer, and enhancing the pleasant multimedia H2M/M2H intuitive interactions. It hides "behind the curtains" the mental and cognitive overloads by: shifting the most portion of ICT-related interactions from H2H to M2M zone; using AR pathways for delivering status information and advices for elderly end users; increasing AR-readiness of the available ICTs for AR-output of data for non-obtrusive H2M/M2H interactions, and improving every-day communication and service needs. It could be the integral platform and paradigm for overcoming cognitive, cultural, mental, gender/ethical, institutional, regional, and national barriers and enabling the targeted delivery of integrated elderly care on European and worldwide scale by joining efforts across governmental, non-governmental, and volunteer elderly care organizations and individuals. In this way elimination of any kinds of "borders" between people at European (and worldwide) scale by targeted efforts can strengthen the relationships between the different age categories of people and various elderly communities despite their intrinsic or imposed differences.

## Acknowledgment

The work was partially supported by Ukraine-France Collaboration Project (Programme PHC DNIPRO) (http://www.campusfrance.org/fr/dnipro), EU TEMPUS LeAGUe project (http://tempusleague.eu), and Croatian Centre of Research Excellence for Data Science and Advanced Cooperative Systems.